\documentclass[doublecol]{epl2}

\usepackage{amssymb}

\usepackage{graphicx}
\title{Intrinsic Percolative Superconductivity in K$_x$Fe$_{2-y}$Se$_2$ Single Crystals}
\shorttitle{Intrinsic Percolative Superconductivity in K$_x$Fe$_{2-y}$Se$_2$ Single Crystals} 

\author{B. Shen\inst{1} \and B. Zeng\inst{1} \and G. F. Chen\inst{2} \and J. B. He\inst{2}\and D. M. Wang\inst{2} \and H.
Yang\inst{3} \and H. H. Wen\inst{1,3 (a)} } \shortauthor{B. Shen
\etal}
 \institute{
  \inst{1} Institute of
Physics and Beijing National Laboratory for Condensed Matter
Physics, Chinese Academy of Sciences, P.O. Box 603, Beijing
100190, People's Republic of China\\
  \inst{2} Department of Physics, Renmin University of China,
Beijing 100872, China\\
  \inst{3} National Laboratory of Solid State Microstructures
and Department of Physics, Nanjing University, Nanjing 210093,
China} \pacs{74.25.Rp}{Pairing symmetries (other than s-wave) }
\pacs{74.70.Dd}{Ternary, quaternary, and multinary compounds
(including Chevrel phases, borocarbides, etc.) }
\pacs{74.25.Ha}{Magnetic properties including vortex structures
and related phenomena (for vortices, magnetic bubbles, and
magnetic domain structure }

\abstract{ Magnetic field penetration and magnetization hysteresis
loops (MHLs) have been measured in K$_x$Fe$_{2-y}$Se$_2$ single
crystals. The magnetic field penetration shows a two-step feature
with a very small full-magnetic-penetration field ($\approx$ 300
Oe at 2 K), and accordingly the MHL exhibits an abnormal vanishing
of the central peak near zero field below 13 K. The width of the
MHL in K$_x$Fe$_{2-y}$Se$_2$ at the same temperature is in general
much smaller than that measured in the relatives
Ba$_{0.6}$K$_{0.4}$Fe$_2$As$_2$ and
Ba(Fe$_{0.92}$Co$_{0.08}$)$_2$As$_2$, and the MHLs in the latter
two samples show the normal central peak near zero field. All
these anomalies found in K$_x$Fe$_{2-y}$Se$_2$ can be understood
in the picture that the sample is percolative with weakly coupled
superconducting islands.}

\begin{document}

\maketitle

\section{Introduction}

The high temperature superconductivity discovered in the iron
pnictides is a surprise since they contain iron elements which are
normally believed to have strong magnetic moment and thus
detrimental to superconductivity. The important issue concerning
the superconductivity is the pairing mechanism. Experimentally it
was found that the superconducting state is at the vicinity of a
long range antiferromagnetic (AF) order\cite{DaiPC}, and the
superconducting state is recovered when the AF state is
suppressed. Further experiments have proved that the AF spin
fluctuation\cite{ImaiNMR} and the multi-band effect\cite{FangLei}
are two key factors for driving the system into superconductive.
These give partial support, although not the complete, to the
picture that the pairing may be established via inter-pocket
scattering of electrons between the hole pockets (around $\Gamma$
point) and electron pockets (around M point), leading to the
pairing manner of an isotropic gap on each pocket but with
opposite signs between them (the so-called
S$^\pm$).\cite{Mazin,Kuroki} Recently a new Fe-based
superconducting system $A_xFe_{2-y}Se_2$ (A= alkaline metals,
x$\leq$1, y$\leq0.5$) were discovered with the transition
temperature above 30 K.\cite{ChenXL} The interests to this system
have been stimulated quickly since both the band structure
calculations\cite{BandCal,LuZY} and the preliminary angle resolved
photo-emission spectroscopy (ARPES)
measurements\cite{FengDL,ZhouXJ,DingH} indicate that the band near
the $\Gamma$-point seems diving far below the Fermi energy,
leading to the absence of the hole pockets which are prerequisites
for the pairing model mentioned above. On the other hand, based on
the simple charge counting, it was speculated that there must be
Fe vacancies in the sample,\cite{FangMH} which may or may not
order in the superconducting state.\cite{BaoW,HanF} If the Fe
vacancies are ordered, band structure calculations would predict
that\cite{BandCal} there is probably a small band gap, which may
interpret the insulating behavior of the so-called parent phase.
Some preliminary experiments using transmission electron
microscopy (TEM)\cite{LiJQ}indicate that the Fe vacancies indeed
form some kind of orders. Given the easy mobility of the Fe
vacancies and probably also the potassium, cautions are required
to draw any conclusions about the ordering form of Fe vacancies,
especially when the samples are needed to be pre-treated for some
measurements. Due to the uncertainties about the ordering of Fe
vacancies, it is curious to know whether the superconducting state
is uniform. In this Letter we report the measurements on the
penetration of magnetic field into the single crystal sample. Our
results clearly illustrate an abnormal magnetic field penetration
in the single crystals, which can be easily explained as due to
the percolative superconductivity.

\section{Sample preparation and experiment}
The K$_x$Fe$_{2-y}$Se$_2$ single crystals were synthesized by the
Bridgeman method\cite{ChenGFKFeSe}. The typical dimensions of the
samples for the magnetization measurements were
2$\times$2$\times$0.5 mm$^3$. The magnetization measurements were
done with the vibrating sample magnetometer based on the Quantum
Design instrument physical property measurement system (PPMS) with
the temperature down to 2 K and magnetic field up to 9 T. For
investigating the magnetic field penetration, we have used a
special power supply for the magnet, which can sweep the magnetic
field in a rate as low as 0.4 Oe/s and the data were collected
densely during the field sweeping process. All the magnetization
measurements were done with the magnetic field parallel to c-axis.
The resistive measurements were also measured leading to the
determination of the upper magnetic fields H$_{c2}^c$ (H$\|$c).
The dc magnetization measurements were done with a superconducting
quantum interference device (SQUID) of Quantum Design.

\section{Results and discussion}
The temperature dependence of dc magnetization at H = 20 Oe for
K$_x$Fe$_{2-y}$Se$_2$ and the magnetic hysteresis loops (MHL) for
different iron-based superconductors are presented in Fig. 1. The
large difference between zero-field cooling (ZFC) and
field-cooling (FC) magnetizations indicates a strong magnetization
hysteresis in the sample at a low field. A rough estimate on the
diamagnetic signal indicates that the Meissner screening at H = 20
Oe is about 100 $\%$, this however does not suggest that the
sample has a full superconducting volume. We will further
illustrate this point below. The MHL measured at 2 K is presented
in Fig. 1(b). One can easily see an abnormal minimum of
magnetization near the zero field. From an enlarged view in the
low field region during the magnetic penetration, as shown in Fig.
1(c), a feature of two-step penetration with a very small
full-magnetic-penetration field ($\approx$ 300 Oe at 2 K) is found
in the initial part of the magnetic penetration curve which is
corresponding to the abnormal minimum magnetization, the dip, in
MHL when the field is reduced back to zero. After the initial
dropping down at H$_{p1}$, the magnetization rises up again and
reaches a maximum at about H$_{p2}$ = 0.5 T, as shown in Fig. 1(b)
and (c). This unique two-step magnetic penetration is abnormal in
a common sense to all other uniform type-II superconductors, but
was observed in some overdoped cuprate superconductors in which
the phase separation may occur.\cite{WenEPL} In other iron-based
superconductors such as Ba$_{1-x}$K$_{x}$Fe$_2$As$_2$ (BaK122),
Ba(Fe$_{1-x}$Co$_x$)$_2$As$_2$ (BaFeCo122) and so on, a sharp
magnetization peak can be observed near zero field (shown in Fig.
1(d)),\cite{YangHAPL,ShenBPRB} and just one magnetic penetration
peak near about H = 0.5 T can be found in the MHL. In additional
to the abnormal minimum of magnetization near zero field and the
two-step penetration on the MHL curve below about 13 K, when
compared to BaK122 and BaFeCo122, the width of the MHL for
K$_x$Fe$_{2-y}$Se$_2$ at the same temperature is about 50 to 100
times smaller, which suggests a low superconducting critical
current density in the present system.

In the superconducting state of a uniform type-II superconductor,
the condensate of the superconducting electrons will expel the
external magnetic field. When the external field H is higher than
the lower critical field H$_{c1}$ at the edge, the quantized
magnetic vortices will be formed and penetrate into the interior
of the sample. The spatial distribution of the density of these
vortices, also called as flux profile, is illustrated in the inset
of Fig. 2(b). The curve of M vs. H will deviate from the linear
relation M = -H/4$\pi$ (Meissner screening) at H$_{c1}$ for a
cylinder sample (assuming that the Bean-Livingston barrier and the
geometrical barrier of the flux penetration is negligible). The
magnetization M will continue to grow until the flux fronts of
both side meet at the center of the sample (H = H$_p$). By further
increasing the external field, more magnetic vortices will creep
into the sample, and the magnetization will start to drop because
the slope of dH/dx becomes smaller. Therefore, for any uniform
superconductor, a central magnetization peak will appear near zero
field due to the penetration of magnetic flux. Based on the simple
Bean critical state model, one can estimate the full penetration
magnetic field as $dH/dx=\mu_0j_c$ with $j_c$ the critical current
density. Taking $j_c$ = 10$^{5} A/cm^2$ and based on the Bean
critical state model, we have $H_p\approx 1.26 T $ across a sample
area of 1 mm. Normally this value will be reduced because of the
demagnetization effect. As an example, in Fig. 1 (d) we show MHLs
measured at 2 K for optimally doped BaK122 and BaFeCo122 single
crystals. A clear penetration peak can be observed at about 0.5 T
here for both samples. When the field is reduced down from a high
value (from both the positive and the negative side), a central
magnetization peak will appear near zero field due to the
establishing of a large superconducting current.

\begin{figure}
\includegraphics[width=8cm]{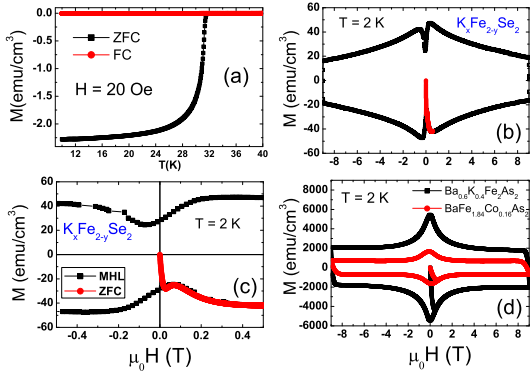}

\caption {(color online) (a) The temperature dependence of
magnetization measured at H = 20 Oe with the field-cooling and
zero-field-cooling processes. (b) The MHL measured at T = 2 K and
(c) An enlarged view in the low field region during the magnetic
field penetration. The dark squares represent the data measured
from 9 T to -9 T and back to 9 T at a field sweeping rate of 200
Oe/s. The red circles show the data in the low field region
measured with a rate of 5 Oe/s. An abnormal dip of MHL appears
near zero field. (d) The MHLs measured at T = 2 K for
Ba$_{1-x}$K$_{x}$Fe$_2$As$_2$ and Ba(Fe$_{1-x}$Co$_x$)$_2$As$_2$.
Clearly a maximum appears near zero field for both samples, in
contrast with that in the samples K$_x$Fe$_{2-y}$Se$_2$.}
\label{fig1}
\end{figure}

\begin{figure}
\includegraphics[width=8cm]{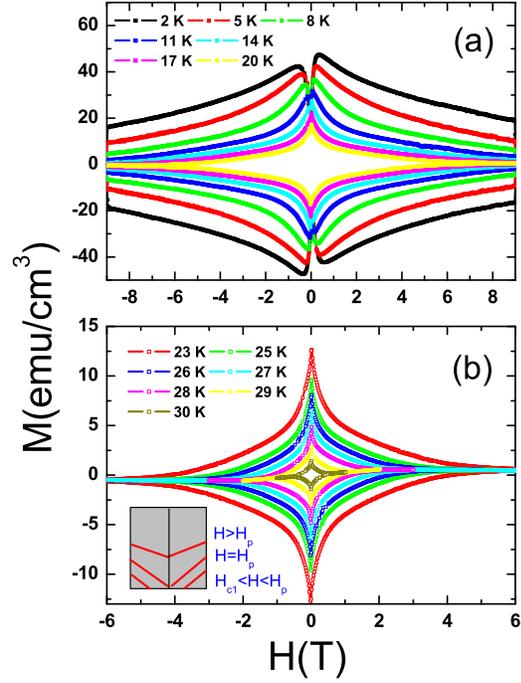}
\caption {(color online) The MHLs measured with the magnetic field
sweeping field of 200 Oe/s at (a) 2 - 20 K and (b) 23 - 30 K.
Below about 13 K, an abnormal dip, instead of a peak appears near
the zero field. Above 13 K, a normal central peak appears near
zero field. The inset in (b) shows the magnetic flux profile when
the external field is smaller, equal to and higher than the full
magnetic penetration field H$_p$ at which the flux fronts meet at
the center of the sample.} \label{fig2}
\end{figure}

Magnetic hysteresis loops of the K$_x$Fe$_{2-y}$Se$_2$ samples
measured at different temperatures (from 2 K to 30 K) are
presented in Fig. 2. When the temperature is below about 13 K, the
MHLs exhibit the abnormal magnetic penetration effect, with a
two-step structure. A similar phenomena was observed in the same
system with the samples synthesized in a different
way.\cite{MaYW}. While at high temperatures, the MHL becomes
"normal" with only one central peak, instead of the dip near zero
magnetic field. As mentioned above, in other iron-based
superconductors, such as optimally doped BaK122 and BaFeCo122, a
sharp central magnetization peak was observed near zero field even
at the lowest temperature. This central peak can be understood in
the following way: when the external field is swept back to zero,
because of the small absolute value of H(x) near the edge, the
slope of H(x) and thus the critical current density near the edge
is much larger than that in the interior part and then a much
enhanced magnetization will appear. In order to investigate the
abnormal dip near zero field, we measured the magnetic penetration
from 2 K to 20 K in a very detailed way with a very slow magnetic
field sweeping rate (dH/dt = 5 Oe/s). Fig.3 shows the
magnetization vs. field curves in the initial penetration period.
The two-step feature with a small full penetration can be observed
below 13 K. The first-step magnetic penetration field (H$_{p1}$)
is very small($\approx$ 300 Oe at 2 K), which is corresponding to
the abnormal dip of the MHL when the field is swept continuously
through the zero field. This first penetration peak is followed by
a second one at about 0.5 T at 2 K. With the temperature
increasing, the first-step penetration magnetization decreases
gradually while the position of H$_{p1}$ changes little. The
second-step penetration magnetic field($H_{p2}$) shifts to a lower
field with the temperature increasing and finally converge with
the first-step penetration field on the MHL curve measured at 13
K.

\begin{figure}
\includegraphics[width=8cm]{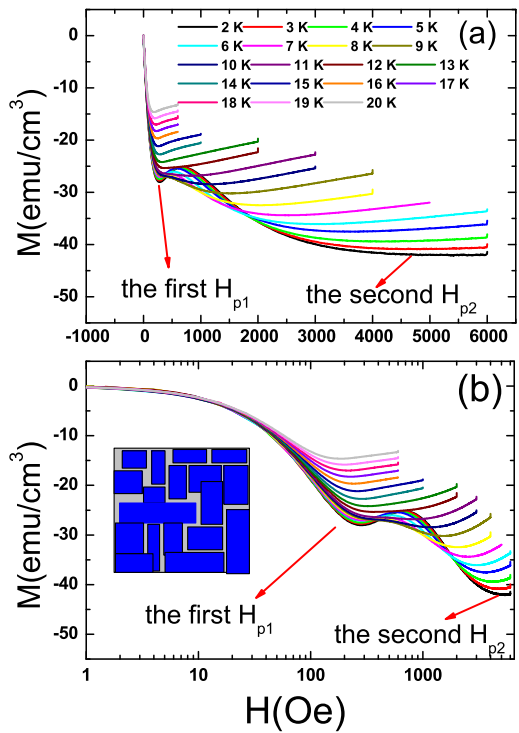}
\caption {(color online) (a) The magnetization versus magnetic
field measured in the low field region with the magnetic field
sweeping rate of about 5 Oe/s. A two-step penetration, as marked
by the full-penetration fields H$_{p1}$ and H$_{p2}$ can be easily
observed below 13 K. (b) The same data illustrated in a
semi-logarithmic way. The inset gives a cartoon picture of the
percolative superconductivity. The blue blocks represent the
superconducting area.} \label{fig3}
\end{figure}

The two-step penetration and the abnormal dip near zero field on
the MHL for K$_x$Fe$_{2-y}$Se$_2$ can be understood with the
picture that the sample is percolative with weakly coupled
superconducting islands. In low field region, the magnetic flux
will penetrate into the center of the sample easily through the
non-superconducting channels and it is difficult to establish a
high critical current density near zero field. So the flux fronts
can quickly meet at the center of the sample with a very weak full
penetration field (H$_{p1} \approx$ 300 Oe). For the same reason,
the central peak of MHL cannot be observed since the large
critical current density cannot be established. When the field is
further increased, the flux which already reaches the center of
the sample will penetrate the individual superconducting islands
again. This leads to the observation of the second penetration
peak at H$_{p2}$. In samples with poorer transitions, we found
that the two-step magnetic transition and the dip of the MHL at
zero field cannot be observed, since the islands are far apart
each other and the magnetic field penetrate the individual grains
directly. We should emphasize that the abnormal magnetic
penetration in our samples cannot be simply attributed to the
second peak (or called as the fish-tail) effect of magnetization
because of the following two reasons. Firstly, the magnetic fields
for the two maxima on the MHL of K$_x$Fe$_{2-y}$Se$_2$ are too
small to compare with those in BaK122 and BaFeCo122. In the latter
two cases, the magnetic penetration field is about 0.5 T and the
second peak of the magnetization locates at a very high field
(above 9 T at 2 K).\cite{YangHAPL,ShenBPRB} Secondly, as mentioned
before, the width of the MHL is about 50 to 100 times smaller in
K$_x$Fe$_{2-y}$Se$_2$ than in optimally doped BaK122 and
BaFeCo122.

\begin{figure}
\includegraphics[width=8cm]{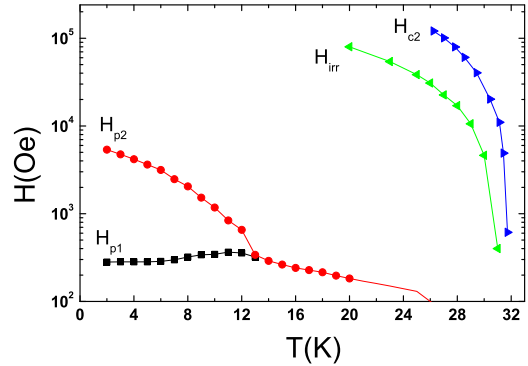}
\caption {(color online) The phase diagram with the first and
second full-magnetic-penetration fields H$_{p1}$ and H$_{p2}$, the
irreversibility field H$_{irr}$ and the upper critical fields
determined from the resistive measurements.} \label{fig4}
\end{figure}

In K$_x$Fe$_{2-y}$Se$_2$ it becomes well known that there are Fe
vacancies. It was proposed that the Fe vacancies may order in
different structures when y = 0.5 or y = 0.4.\cite{FangMH} Some
preliminary experiments using the electron transmission microscope
(TEM) do find the order of Fe vacancies.\cite{LiJQ} Our previous
experiment indicates that the superconductivity can be recovered
from an insulating parent phase just by post-annealing and fast
quenching the sample, this may suggest that the random
distribution of the Fe vacancies help to stabilize the
superconducting phase.\cite{HanF} Regarding the uncertain
structural forms of these Fe vacancies, it is quite possible to
have a phase separated state which contains the superconducting
islands (with randomly distributed  or low density\cite{Canfield}
Fe vacancies) surrounded by the non-superconducting area (with
ordered or high density Fe vacancies). These non-superconducting
area may possess the insulating behavior. Regarding the sharpness
of the magnetic transition and the perfect Meissner screening, our
sample is among the best class reported so far. This allows us to
speculate that almost all the superconducting samples reported so
far in this family may have the same feature of phase separation.
which is agreeable with other measurements
results\cite{DLfeng1,QKxue,Keimer}. the global measurements of
magnetization.

In Fig.4 we show the phase diagram of the K$_x$Fe$_{2-y}$Se$_2$
single crystals. Since the upper critical field H$_{c2}$ and the
irreversibility field H$_{irr}$ are determined by the
superconducting islands here, therefore they can be treated as the
intrinsic properties of the K$_x$Fe$_{2-y}$Se$_2$ system. One can
see that both the H$_{c2}$(T) and H$_{irr}$(T) are very high and
very close to their relatives BaK122 and BaFeCo122. The second
penetration field H$_{p2}$ in our picture corresponds actually to
the magnetic penetration of the superconducting islands. Above
about 13 K, the first penetration and the second one merge, this
is because the weak coupling between the superconducting islands
in the high temperature region becomes very weak against the
magnetic penetration. Our picture naturally explains why the
residual resistivity is large and the normal state resistivity
exhibits a hump-like temperature dependence. All these are induced
by the composed contribution of metallic islands (superconducting
at low-T) and the insulating surrounding areas. A direct proof to
this picture would need a local scanning probe in the
superconductng state.

\section{Conclusion}
In summary, we have measured the magnetic penetration and MHLs in
K$_x$Fe$_{2-y}$Se$_2$ single crystals. An abnormal two-step
magnetic penetration, a dip instead of a peak near zero field on
the MHL and a much reduced magnitude of the MHL were observed. All
these anomalous features can be understood with the phase
separation picture, perhaps electronic in origin. Regarding the
uncertain structural form of the Fe vacancies and their influence
on the electronic properties, we argue that the sample has
percolative superconductivity. This picture recalls local probe
measurements in the superconducting state.

\acknowledgments We appreciate the useful discussions with Takashi
Imai and Minghu Fang. This work is supported by the NSF of China,
the Ministry of Science and Technology of China (973 projects:
2011CBA001000), and Chinese Academy of Sciences.


\begin{thebibliography}{0}
 \bibitem{DaiPC} \Name{De la Cruz C.} \emph{et al.} \REVIEW{Nature}{453} {2008} {899}.
 \bibitem{ImaiNMR}\Name{Ning F. L.} \emph{et al.} \REVIEW{Phys. Rev. Lett} {104} {2010} {037001}.
\bibitem{FangLei} \Name{Fang L.} \emph{et al.} \REVIEW{Phys. Rev. B} {80} {2009} {140508(R)}.
\bibitem{Mazin} \Name{Mazin I. I.} \emph{et al.} \REVIEW{Phys. Rev. Lett.} {101} {2008} {057003}.
\bibitem{Kuroki} \Name{Kuroki K.} \emph{et al.} \REVIEW{Phys. Rev. Lett.} {101} {2008} {087004}.
\bibitem{ChenXL} \Name{Guo J.} \emph{et al.} \REVIEW{Phys. Rev. B} {82} {2010} {180520(R)}.
\bibitem{BandCal}\Name{Nebrasov I. A} \emph{et al.} \REVIEW{JETP Letters} {93} {2011} {166}.
\bibitem{LuZY}\Name{Yan X.-W.}\emph{et al.}arXiv:1012.5536 (2011)
\bibitem{FengDL}\Name{Zhang Y.} \emph{et al.} \REVIEW{Nature Materials} {10} {2011} {273}.
\bibitem{ZhouXJ}\Name{Mou D. X.} \emph{et al.} \REVIEW{Phys. Rev. Lett.} {106} {2011} {107001}.
\bibitem{DingH}\Name{Qian T.} \emph{et al.} \REVIEW{Phys. Rev. Lett.} {106} {2011} {187001}.
\bibitem{FangMH}\Name{Fang M. H.} \emph{et al.} \REVIEW{EPL} {94}
{2011} {27009}.
\bibitem{BaoW}\Name{Bao W.} \emph{et al.}  arXiv:1102.3674 (2011).
\bibitem{HanF} \Name{Han H.} \emph{et al.} arXiv:1103.1347 (2011).
\bibitem{LiJQ}\Name{Li J.Q} \emph{et al.} \REVIEW{Phys. Rev. B} {83} {2011} {140501(R)}.
\bibitem{ChenGFKFeSe}\Name{Wang D. M.} \emph{et al.} \REVIEW{Phys. Rev. B} {83} {2011} {132502(R)}.
\bibitem{WenEPL}\Name{Wen H. H.} \emph{et al.} \REVIEW{EPL} {57} {2002} {260}.
\bibitem{MaYW}\Name{Gao Z. S.} \emph{et al.} arXiv:1103.2904 (2011).
\bibitem{YangHAPL}\Name{Yang H.} \emph{et al.} \REVIEW{APL} {93} {2008} {142506}.
\bibitem{ShenBPRB}\Name{Shen B.} \emph{et al.} \REVIEW{Phys. Rev. B}
{81} {2010} {014503}.
\bibitem{Canfield}\Name{Hu R.} \emph{et al.} \REVIEW{Supercond. Sci. Technol.}
{24} {2011} {065006}.
\bibitem{DLfeng1}\Name{Ye. Z. R.} \emph{et al.} arXiv:1105.5242 (2011).
\bibitem{QKxue}\Name{Li. W.} \emph{et al.}  arXiv:1108.0069 (2011).
\bibitem{Keimer}\Name{Park. J. T.} \emph{et al.}  arXiv:1107.1703 (2011).

\end{thebibliography}
\end{document}